\documentclass[aps,prb,twocolumn,superscriptaddress,showpacs,floatfix,longbibliography]{revtex4-2}
\usepackage{amsmath,amssymb,amsfonts,float,graphics,epsfig,epstopdf,color,verbatim,tabularx,bm,multirow,appendix,wasysym}
\usepackage[utf8]{inputenc}
\usepackage[T1]{fontenc}
\usepackage{cancel}
\usepackage{xcolor}
\usepackage{dsfont}
\usepackage{textcomp}
\usepackage{yfonts}
\usepackage{bm}
\usepackage{subfigure}
\usepackage{amsmath}
\usepackage{graphicx}
\usepackage{verbatim}
\usepackage{hyperref}
\usepackage{multirow}
\usepackage{braket}
\usepackage[normalem]{ulem}
\usepackage{lipsum}
\usepackage{tikz}
\usepackage{pdfsync}
\usetikzlibrary{calc}
\usetikzlibrary{shapes.multipart}

\newcommand{\comments}[1]{}

\newcommand{\furina}[1]{\textcolor{blue}{\textbf{[Furina: ]}}}
\newcommand{\stkout}[1]{\ifmmode\text{\sout{\ensuremath{#1}}}\else\sout{#1}\fi}

\makeatletter
\def\l@subsubsection#1#2{}
\makeatother

\begin{document}

\title{Singularity and universality from von Neumann to R\'enyi entanglement entropy and disorder operator in Motzkin chains}

\author{Jianyu Wang}
\affiliation{State Key Laboratory of Surface Physics and Department of Physics, Fudan University, Shanghai 200438, China}
\affiliation{New Cornerstone Science Laboratory, Department of Physics, School of Science, Westlake University, Hangzhou 310024, China}
\affiliation{Institute for Theoretical Sciences, Westlake University, Hangzhou 310024, China}
\affiliation{Key Laboratory for Quantum Materials of Zhejiang Province, School of Science, Westlake University, Hangzhou 310024, China}
\affiliation{Institute of Natural Sciences, Westlake Institute for Advanced Study, Hangzhou 310024, China}

\author{Zenan Liu}
\affiliation{Department of Physics, School of Science and Research Center for Industries of the Future, Westlake University, Hangzhou 310030,  China}
\affiliation{Institute of Natural Sciences, Westlake Institute for Advanced Study, Hangzhou 310024, China}

\author{Zheng Yan}
\email{zhengyan@westlake.edu.cn}
\affiliation{Department of Physics, School of Science and Research Center for Industries of the Future, Westlake University, Hangzhou 310030,  China}
\affiliation{Institute of Natural Sciences, Westlake Institute for Advanced Study, Hangzhou 310024, China}

\author{Congjun Wu}
\email{wucongjun@westlake.edu.cn}
\affiliation{New Cornerstone Science Laboratory, Department of Physics, School of Science, Westlake University, Hangzhou 310024, China}
\affiliation{Institute for Theoretical Sciences, Westlake University, Hangzhou 310024, China}
\affiliation{Key Laboratory for Quantum Materials of Zhejiang Province, School of Science, Westlake University, Hangzhou 310024, China}
\affiliation{Institute of Natural Sciences, Westlake Institute for Advanced Study, Hangzhou 310024, China}
% *****************************
\begin{abstract}
The R\'enyi entanglement entropy is widely used in studying quantum entanglement properties in strongly correlated systems, whose analytic continuation as the R\'enyi index $n \to 1$ is 
%always 
often believed to yield the von Neumann entanglement entropy. 
However, earlier findings indicate that this process exhibits a singularity for the colored Motzkin spin chain problem, leading to different scaling behaviors of $\sim \sqrt{l}$ and $\sim \log{l}$ for the von Neumann and R\'enyi entropies, respectively. 
Our analytical and numerical calculations confirm this transition, which can be explained by the exponentially increasing density of states in the entanglement spectrum that we extract numerically. 
Disorder operators are further employed under various symmetries to study such a system. 
Both analytical and numerical results demonstrate that the scaling of the disorder operators also follows $\log{l}$ as the leading behavior, matching that of the R\'enyi entropy. 
We propose that the coefficient of the term $\log{l}$ is a universal constant shared by both the R\'enyi entropies and disorder operators. 
This universal constant could potentially help capture the underlying constraint physics of Motzkin walks.
\end{abstract}
% ************************************************************
% ============================================================
\date{\today}
\maketitle

% ============================================================
% ************************************************************
\section{Introduction}
In recent years, the interflow and mutual learning between condensed matter and quantum information have inspired increasingly fruitful research~\cite{Amico2008entanglement,laflorencie2016quantum,zeng2019quantum}. 
In the field of quantum information, the entanglement entropy (EE) plays a key role in measuring information and chaos~\cite{nielsen2010quantum,wilde2013quantum}. 
As one of the basic properties of quantum mechanics, quantum entanglement itself is difficult to measure. Moreover, the introduction of EE in condensed matter physics has revealed rich physics, such as topological entanglement entropy and long-range entanglement in highly entangled matter~\cite{Kitaev2006,Levin2006}.
One important topic in condensed matter physics is using EE to probe the intrinsic physics of many-body systems~\cite{vidal2003entanglement,Korepin2004universality,Kitaev2006,Levin2006,wang2024probing}. 
Among many intriguing features, EE offers a direct connection between the conformal field theory (CFT) and categorical description of the problems beyond traditional observables~\cite{Calabrese2008entangle,Fradkin2006entangle,Nussinov2006,Nussinov2009,CASINI2007,JiPRR2019,ji2019categorical,kong2020algebraic,XCWu2020,ding2008block,Tang2020critical,JRZhao2020,XCWu2021,JRZhao2021,BBChen2022,JRZhao2022,zyan2021entanglement}. 

Unlike in few-body systems, the scaling behaviors of EE in quantum many-body systems reveal universal properties, such as the central charge~\cite{Calabrese_2004,Calabrese2008entangle,ding2024trackingvariationentanglementrenyi,ding2024reweightannealing}, number of Goldstone modes~\cite{metlitski2011entanglement,deng2023improved,deng2024diagnosing,wang2024ee}, and quantum dimension of topological order~\cite{Kitaev2006,Levin2006,Grover2013EE-QSL,JRZhao2021}.
Within the framework of CFT~\cite{nozaki2014quantum,cardy2016entanglement,bueno2015universality,CASINI2007,casini2012positivity},
the EE with cornered cuttings in two-dimensional quantum systems usually obeys the area-law  $s=al+b\ln{l}+\delta$, where $l$ is the length of the entangled boundary; $b$ is related to the angles of corners; $a$ is generally thought to be UV dependent. 
Meanwhile, the coefficients $b$ and $\delta$ usually can be employed to extract universal information, and detect novel phases and criticalities~\cite{laflorencie2016quantum}. 

The von Neumann (VN) EE $s^{\text{VN}}_A=-\mathrm{Tr} (\rho_A\log \rho_A)$ ($\rho_A$ is the reduced density matrix) as the generalization of the Shannon entropy in quantum mechanics is widely used in studying the above questions~\cite{longo2021neumann,boes2019neumann,giraldi2001quantum,donnelly2012decomposition}. 
However, due to the difficulty in the calculation of the VN EE, the R\'enyi EE is much more commonly used both in field theory and numeric calculations. 
The definition of the R\'enyi EE is $s^{(n)}_{A}=\frac{1}{1-n}\log[\mathrm{Tr}(\rho_A^n)]$ with $n$ the R\'enyi index. 
Formally, the R\'enyi EE becomes
the VN EE as $n \rightarrow 1$. 
Although the R\'enyi EE loses certain properties including additivity and sub-additivity~\cite{cubitt2008counterexamples,PhysRevB.99.045155,collins2011random,d2021alternative,fuentes2022renyi}, it is strongly believed to yield the same scaling behaviors as the VN EE, such as the area and volume laws. 
It has been demonstrated in the past that all the coefficients of the VN EE for each $O(l)$ term can be obtained via the analytic continuation of $n$. 
For a large number of examples including both field theoretical and numerical results, this common belief has been massively tested~\cite{Eisert2010ColloquiumAreaLaws,Flammia2009TopologicalEntanglementRenyi}.

However, the colored Motzkin spin chain was found as a counter example 
in previous studies
~\cite{RM2016, Sugino:2018ab, Menon:2024vic}.
There is a singularity in the limit of R\'enyi index $n\rightarrow 1$, in which the coefficient of the leading term $\log l$ will be divergent. It actually indicates that an extra term $\sqrt{l}$ becomes the leading term in the VN EE. This result provides a counter-example for the analytic continuation from R\'enyi to von Neumann EE, which cautions people should be careful in EE studies.

On the other hand, the disorder operator (DO) is a non-local observable, similar to the EE, which is proposed to extract the high-form symmetry and CFT information of quantum many-body systems~\cite{Wu2021,lake2018higherform,fradkin2017disorder,liu2024measuring}. 
It has been successfully used to detect the high-form symmetry breaking at the Ising transition~\cite{zhao2021}. 
The current central charges in CFT can be captured from the DO at phase transitions of the $O(2)$ and $O(3)$ universal classes in (2+1)D\cite{Wang2021,Wang2022}. 
DOs are also designed for fermionic systems to explore the universal feature of the Fermi liquid, Luttinger liquid, and quantum critical point (QCP) in fermionic systems\cite{jiang2023versus,LiuF2023,liu2023disorder}. 
The DO satisfies the universal scaling behaviors, where the logarithmic term usually reflects the general feature of CFT at a conformal invariant QCP.

In this article, we systematically study the scaling behaviors of VN and R\'enyi EE of a 1D colored Motzkin spin chain with different spin $S$ both analytically and numerically. 
The scaling behaviors of DO for the Motzkin spin are also investigated, which are similar to the R\'enyi EE. 
Moreover, the coefficient of the logarithmic term is universal for any spin value, regardless of DO or EE. 
%It can help us identify the system of Motzkin walks.

% \jerry{add structure}

The rest part of this paper is organized as follows. 
In Sec. II, the background of the Motzkin chain is presented.  
In Sec. III, the analytic solution of EE is given, encompassing an analysis of R\'enyi entanglement entropy to facilitate further discussions. 
In Sec. IV, we introduce the algorithm for Monte Carlo simulations and provide a detailed discussion of the numerical results for entanglement entropy. 
In Sec. V, we focus on the disorder operators of the Motzkin chain, exploring both analytical and numerical perspectives, and reveal that these disorder operators scale as \(\frac{3}{2}\log{l}\). 
The summary and conclusion are presented in Sec. V.

% ==========================

\section{Background of Motzkin chains}
% \jerry{add section summary}
In this section, we review the concept of Motzkin chain, which is a 1D spin chain inspired by the Motzkin number~\cite{Motzkin1948RelationsHypersurfaceCross,Donaghey1977MotzkinNumbers} from the combinatorial mathematics.
%\zyan{do we have any reference here?}\jerry{added}
% The concept of Motzkin walk and Motzkin spin chain model is reviewed in this section.

\begin{figure}[t]
  \centering\includegraphics[width=.7\linewidth]{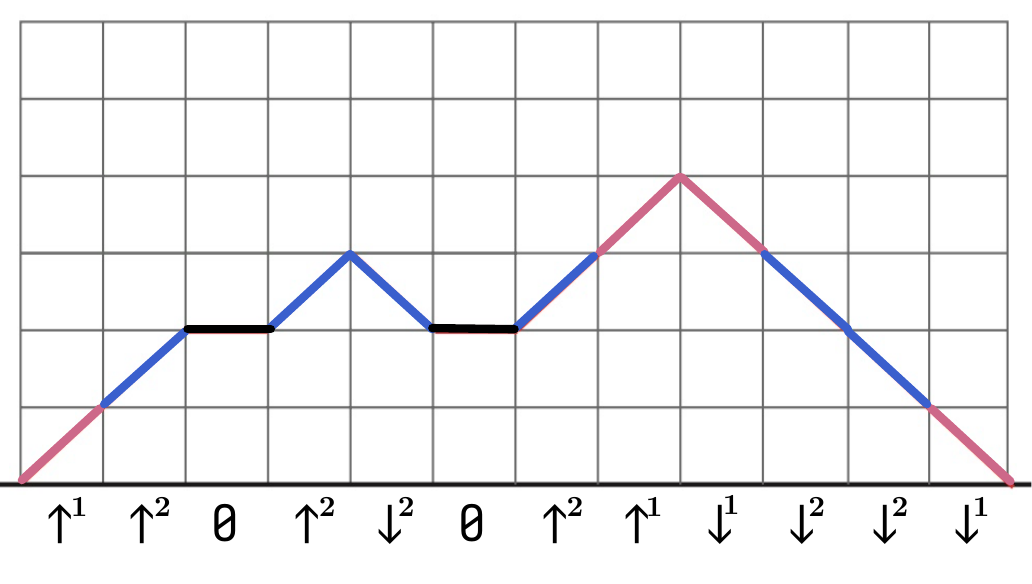}
    \caption{\label{fig:walk-example}Example of a $2$-colored Motzkin walk (colored path). The walk starts and ends on the $x$-axis, and cannot cross below it. It can be mapped to a $S^z$ configuration of a spin-2 chain. In this figure, the pink upward and downward steps are mapped to $S^z=\pm 1$, the blue steps are mapped to $S^z=\pm 2$, and the black horizontal steps are mapped to $S^z=0$.}
\end{figure}

Motzkin walk or Motzkin path is a kind of non-negative lattice path defined as follows. 
Considering random walks starting from the point $(0,0)$ and ending at $(2l,0)$ on the $x$-$y$ plane as shown in Fig.\ref{fig:walk-example}, they consist of three types of steps: upward steps $\diagup$, downward steps $\diagdown$, and horizontal steps $\substack{\\[-0.7ex] \overline{\quad} ~}$, represented as vectors $(1,1)$, $(1,-1)$ and $(1,0)$, respectively. 
Then Motzkin walks are random walks that do not cross below the $x$-axis ($y=0$). 
Bravyi et al.~\cite{Bravyi:2012aa} introduced a frustrated-free spin-1 chain model whose unique ground state is the superposition of all Motzkin walks states, achieved by mapping the spin $S^z$ states $\{ \uparrow,0,\downarrow\}$ to the steps $\{\diagup, \substack{\\[-0.7ex] ~\overline{\quad} ~}, \diagdown\}$, respectively. Movassagh and Shor~\cite{RM2016} then generalized this model into an $S$-colored Motzkin model, where the upward or downward steps are colored using $S$ different colors. 
This model can be translated to a spin-$S$ quantum chain. 
An example of the $2$-colored Motzkin walk is shown in Fig.~\ref{fig:walk-example}. 
For a 1D $S$-colored Motzkin model of size $2l$, the Hamiltonian is constructed by local projection operators with the following form:
\begin{equation}
\label{eq:hamiltonian}
    H= \Pi^\text{boundary}  + \sum_{j=1}^{2l-1} \Pi_{j,j+1}^\text{cross} + \sum_{j=1}^{2l-1} \Pi_{j,j+1}^\text{exchange},
\end{equation}
with
\begin{equation}
    \begin{split}
        &\Pi^\text{boundary} = \sum_{k=1}^{S} \left(  \ket{\downarrow^{k}}_{1,1}\bra{\downarrow^{k}} + \ket{\uparrow^{k}}_{2l,2l}\bra{\uparrow^{k}} \right),\\
        &\Pi_{j,j+1}^{\text{cross}} \hspace{1.3em}= \sum_{k\ne k'}^S \ket{\uparrow^{k}\downarrow^{k'}}_{j,j+1} \bra{\uparrow^{k}\downarrow^{k'}},\\
        &\Pi_{j,j+1}^{\text{exchange}} = \sum_{k=1}^{S} \ket{D^k}_{j,j+1}\bra{D^k} + \ket{U^k}_{j,j+1}\bra{U^k}\\
        & \hspace{5em} + \ket{V^k}_{j,j+1}\bra{V^k},
    \end{split}
\end{equation}
where the superscript $k$ indicate colors, {\it i.e.}, $\uparrow^k$ stands for the $S^z=k$ state and $\downarrow^k$ stands for the $S^z=-k$ state. 

The boundary term ensures that the first and last steps do not pass below the $x$-axis. 
The cross term keeps the correct order of different color pairs. 
As a counter-example, $\uparrow^{k} \uparrow^{k'} \downarrow^{k} \downarrow^{k'}$ is not allowed if $k\ne k'$. 
And the  exchange term contains states $\ket{D^k} =  \ket{\downarrow^{k}0} -  \ket{0 \downarrow^{k}}$, $\ket{U^k}= \ket{\uparrow^{k}0}-  \ket{0 \uparrow^{k}}$ and $\ket{V^k} = \ket{\uparrow^{k}\downarrow^{k}}-  \ket{0 0}$, which leads to the steps exchanging of 
\begin{equation}
        \diagup \substack{\overline{\quad} \\[0.5ex]~} \leftrightarrow \substack{\underline{\quad} \\[-2.8ex]~ }\diagup\,, \qquad
        \diagdown \substack{\underline{\quad} \\[-2.8ex]~ } \leftrightarrow  \substack{\overline{\quad} \\[0.5ex]~}\diagdown\,, \qquad
        \diagup \diagdown  \leftrightarrow \substack{\underline{\quad} \\[-2.8ex]~ }\substack{\underline{\quad} \\[-2.8ex]~ }\,.
\end{equation}
Notice that there is no such step exchange as 
\begin{equation}
    \diagdown \diagup  \leftrightarrow \substack{\overline{\quad} \\[0.5ex]~} \substack{\overline{\quad} \\[0.5ex]~}\,.
\end{equation}
This property ensures that the walk height remains positive. 

The ground state of the Hamiltonian~\eqref{eq:hamiltonian} is the superposition of all the Motzkin walk configurations $\ket{M_i}$\cite{Bravyi:2012aa},
% \zyan{is there any reference?}\jerry{added}
\begin{equation}
\label{eq:groundstate}
    \ket{G} = \frac{1}{\sqrt{N}}\sum_i^N \ket{M_i},
\end{equation}
where $N$ is the number of all the Motzkin walks allowed. 

%------------------------
%---------------------------

%%%%%%%%%%%%%%%%%%%%%%%%%%%%%%%
\section{The VN and R\'enyi EE's}
The entanglement between two subsystems by cutting the Motzkin chain at the midpoint is analytically studied in this section. 
It is worth noting that the similar results have been studied in Refs.~\cite{RM2016, Sugino:2018ab}, 
{\it i.e.}, the VN EE and R\'enyi EE of colored Motzkin chain exhibit different scalings. 
In order for the article to be self-contained,  we present the calculation of the R\'enyi EE here. 
Our analytical method is different from those in literature.
%which is equivalent to verifying the previous conclusion. 

Consider a $S$-colored Motzkin chain with the size $2l$.
The VN EE of the half-cutting 
% is studied in the following.
% The EE $s_l$ of half-cutting on a $S$-colored Motzkin chain with size $2l$ is studied analytically in this section. 
% The EE of colored Motzkin chain is given in this section. We are interested in the EE $s_l$ of half-cutting on a $S$-colored Motzkin chain with size length $2l$.
% The VN EE ($n\to1$) of a $S$-colored Motzkin chain 
takes the form as\cite{RM2016}
\begin{equation}
  \label{eq:vn_entropy}
    s^\text{VN} = a \sqrt{l} + b \log{l} + \text{const.},
\end{equation}
with
\begin{equation}
 \begin{split}
    a =  2  \sqrt{\frac{2 \sqrt{S}}{(2 \sqrt{S} +1) \pi}} \log{S},  \ \ \,
    b = \frac{1}{2}.
  \end{split}
\end{equation}
The R\'enyi EE ($n>1$) is\cite{Sugino:2018ab}
\begin{equation}
  \label{eq:ry_entropy}
  \begin{split}
    s^{(n)} =   b \log{l} + \text{const}.,  
  \end{split}
\end{equation}
with
 \begin{equation}   
  \begin{split}
    b= \frac{3}{2} \left( 1+ \frac{1}{n-1} \right).
  \end{split}
\end{equation}
The VN and R\'enyi EE's exhibit different scaling behaviors.
For comparison, when $S=1$ (colorless case) the leading terms of both the VN EE and R\'enyi EE are $\frac{1}{2}\log{l}$. 
As a notation, the ``$\log$'' in this article stands for the natural logarithm.

For the completeness of the article and convenience of further analyses, we provide the derivation of the R\'enyi
EE $s^{(n)}$. 
This analysis below elucidates the emergence of a singularity at $n\to1$, whereby the VN EE and R\'enyi EE exhibit distinct scaling regimes governed by discontinuous analytic continuations.

% The ground state of a $S$-colored Motzkin chain has been previously described by equation~\eqref{eq:groundstate}. 

\subsection{Analytical calculation of R\'enyi EE}
Given the ground state wavefunction of Eq.~\eqref{eq:groundstate}, 
the eigenvalues of  
the reduced density matrix's  of the half-chain are~\cite{RM2016}
\begin{equation}
  \label{eq:egvl}
\lambda_m \simeq \frac{S^{-m}}{T} \frac{m^2}{l} \exp{\left[-\frac{1}{2} (2+\frac{1}{\sqrt{S}})\frac{m^2}{l}\right]},
\end{equation}
where the index $m$ takes integer values from $0$ to $l$, corresponding to the midpoint height of the Motzkin walk. 
The degeneracy of the eigenvalue $\lambda_m$ is $S^m$, and $\lambda_m$ reaches its maximum at an intermediate value of $m$. 
Additionally, $T$ is a normalization factor that ensures $
\sum_{m} S^m \lambda_m = 1$.

% where the index $m=0,1,\dots,l$, and they have the meaning of the midpoint height of the Motzkin walk. 
% % The eigenvalue $\lambda_m$ take the maximum value at some finite $0<m<l$.
% The degeneracy of eigenvalue $\lambda_m$ is $S^m$, and $\lambda_m$ takes the maximum value at some finite $m$. $T$ is a normalized factor 
% such that $\sum_{m} S^m\lambda_m = 1$.

The $n$-th R\'enyi EE is then given by
\begin{equation}
% \label{eq:ryee-01}
\begin{split}
  s^{(n)} &= \frac{1}{1-n} \log \sum_{m=0}^l  S^m\lambda_m^{n}\\
  &= \frac{1}{1-n} \log \sum_{m=0}^l \frac{S^{(1-n)m}}{T} \frac{m^2}{l} \exp{\left[-\frac{1}{2} (2+\frac{1}{\sqrt{S}})\frac{m^2}{l}\right]} \\
  &= \frac{1}{1-n} \log \sum_{\xi=0}^{\sqrt{l}} \frac{S^{(1-n)\xi\sqrt{l}}}{T} \xi^2 \exp{\left[-\frac{1}{2} (2+\frac{1}{\sqrt{S}})\xi^2\sqrt{l}\right]},
\end{split}
\label{eq:renyi}
\end{equation}
where we set $\xi = m/\sqrt{l}$ in the last line.

% Substituting $m$ with $\xi = m/\sqrt{l}$, 
% \begin{equation}
%     s^{(n)} = \frac{1}{1-n} \log \sum_{m=0}^l 
% \end{equation}

% and using the integral approximation, 

Using the integral approximation to Eq. (\ref{eq:renyi}), the R\'enyi EE ($n>1$) is expressed as 
\begin{equation}
\label{eq:ryee-02}
\begin{split}
s^{(n)} &\simeq \frac{1}{2} \log{l} +   \frac{1}{1-n} \log \tilde{T}^{-n} \int_0^{\sqrt{l}}d\xi \, \xi^{2n} \times\\
        & \exp\Big[  -\frac{n}{2} (2+\frac{1}{\sqrt{S}}) \xi^2  - (n-1)\log{S} \sqrt{l}\, \xi \Big],
\end{split}
\end{equation}
where
\begin{equation}
    \label{eq:ryee-ttilde}
    \tilde{T} =\frac{T}{\sqrt{l}} \simeq \int_0^{\sqrt{l}}  d\xi \xi^2  \exp{\left[-\frac{1}{2} (2+\frac{1}{\sqrt{S}}) \xi^2 \right]}.
\end{equation}
For the case $n>1$, the R\'enyi EE can be simplified to
\begin{equation}
    \label{eq:ryee-n}
    s^{(n)} = \frac{1}{2} \log{l} - \frac{n}{1-n} \log{\tilde{T}} + R,
\end{equation}
where
\begin{equation}
    \label{eq:ryee-s3}
  \begin{split}
     R&=\frac{1}{1-n} \log \int_0^{\sqrt{l}}d\xi \, \xi^{2n} \exp \Big[ -\frac{n}{2} (2+\frac{1}{\sqrt{S}}) \xi^2  \\
    & - (n-1)\log{(S)} \sqrt{l}\, \xi \Big].
    \end{split}
\end{equation}

For convenience, the following functions are defined:
\begin{itemize}
    \item Define functions $p_1$ and $p_2$ as
    \begin{equation}
        \begin{split}
            p_1(n,S) &= \frac{n}{2} (2+\frac{1}{\sqrt{S}}),\\
            p_2(n,S) &= (n-1)\log{S}.
        \end{split}
    \end{equation}
    \item Define function $f_{\alpha}$ as
      \begin{equation}
        \label{eq:rlt-3}
        f_{\alpha}(l) =  l^{\alpha/2}  \exp{ \left[-(p_1 + p_{2}) l \right]}.
      \end{equation}
    \item Define function $F_{\alpha}$ as 
      \begin{equation}
        \label{eq:rlt-2}
        \begin{split}
            &F_\alpha (p_1, p_2, l) = \\
            & \hspace{1em} \int_0^{\sqrt{l}} d \xi  \xi^{\alpha}  \exp{ \left[-p_1(n,S) \xi^2  -p_{2}(n,S) \sqrt{l}\, \xi \right]}.
        \end{split}
    \end{equation} 
\end{itemize}
% the function $\tilde{T}$ in~\eqref{eq:ryee-ttilde} can be expressed as 
% \begin{equation}
%     \tilde{T} = F_2\big(p_1(1,S),0, l\big),
% \end{equation}
% and 
Consequently, $R$ in Eq.~\eqref{eq:ryee-s3} is expressed as 
\begin{equation}
    R = \frac{1}{1-n} \log F_{2n}.
\end{equation}
The positive function $F_\alpha$ has the limit of $\lim_{l \to \infty} F_\alpha \to 0$. Additionally, 
% we can easily establish the following relations:
% \begin{equation}
%         \label{eq:ga-derivative}
%         \frac{d}{dl} F_\alpha = - \frac{p_2 }{2 \sqrt{l}}  F_{\alpha+1} + \frac{1}{2\sqrt{l}} f_{\alpha}(l),
% \end{equation}
% \begin{equation}
%     \label{eq:ga-iteration}
%       (\alpha+1)  F_\alpha = p_2 \sqrt{l} F_{\alpha+1} + 2p_1 F_{\alpha+2} + f_{\alpha}(l).
% \end{equation}
% Rewrite it into 
% \begin{equation}
%      \frac{F_\alpha}{F_{\alpha+1}} = \frac{p_2 \sqrt{l}}{\alpha+1} +  \frac{2p_1}{(\alpha+1) } \frac{F_{\alpha+2}}{ F_{\alpha+1}}+ \frac{f_{\alpha}(l)}{(\alpha+1) F_{\alpha+1}}.
% \end{equation}
% because $F_{\alpha}$ and $f$ are the positive, and $p_2$ is also positive
for the condition that $n, S>1$ and sufficiently large $l$, the following limit can be proved 
% according to equation~\eqref{eq:ga-iteration}:
% \begin{equation}
%     \begin{split}
%       &\lim_{l\to \infty} \frac{F_\alpha}{F_{\alpha+1}} > \lim_{l\to \infty} \frac{p_2\sqrt{l}}{\alpha +1} = +\infty,\\
%        &\lim_{l\to\infty} \frac{f_{\alpha}(l)}{F_{\alpha+1}} = 0.
%     \end{split}
% \end{equation}
% Thus for large enough $l$ we have the relation
\begin{equation}
  \label{eq:FaFa1s}
    F_\alpha \simeq \frac{p_2\sqrt{l}}{\alpha +1} F_{\alpha+1}.
\end{equation}

% And then $F_{2n+2}$ must be of a smaller order than $p_2 \sqrt{l} F_{2n+1}$. 

%---------------------
\subsection{The scaling form of R\'enyi EE}
Now we analyze the scaling of R\'enyi EE in Eq.~\eqref{eq:ryee-n}. 
The first term contributes a logarithmic scaling obviously. 
The second term contributes a constant because $\tilde{T}$ has a upper bound in the limit $l\to \infty$, which is 
\begin{equation}
    \label{eq:tildet}
    \lim_{l\to\infty}\tilde{T} = \sqrt{\frac{\pi}{2}} {\left(2+\frac{1}{\sqrt{S}}\right)}^{-\frac{3}{2}}.
\end{equation}
Then only the third term $R$ is needed to be calculated. 
% In the following paragraphs, we will prove that the scaling of term $R$ is logarithmic. The exact form of R\'enyi EE at the large $l$ limit will also be given.

%\subsection{The logarithmic %scaling of $R$}

We aim to demonstrate that there exists a finite limit for $R$ divided by $\log{l}$ as $l\to\infty$:
\begin{equation}
  \label{eq:s3-log}
\lim_{l\to\infty} \frac{R}{\log(l)}.
\end{equation}
Substituting
Eqs.~\eqref{eq:rlt-3} and\eqref{eq:rlt-2} into  Eq.~\eqref{eq:s3-log} and using the L'H\^{o}pital's rule, we obtain:
\begin{equation}
\label{eq:s3-log-lim}
\begin{split}
\lim_{l\to\infty} \frac{R}{\log(l)} %\\
= &\frac{2n +1}{2n-2} \lim_{l\to\infty} \frac{p_2 \sqrt{l} F_{2n+1}}{ p_2 \sqrt{l} F_{2n+1} +  p_1 F_{2n+2}} \\
= &\frac{2n+1}{2n-2},
\end{split}
\end{equation}
for which the relation of Eq.\eqref{eq:FaFa1s} is used.
% Based on the relation in~\eqref{eq:FaFa1s}, we have demonstrated that
% \begin{equation}
%   \label{eq:s3-log-coef}
%   \lim_{l\to\infty} \frac{R}{\log(l)} = \frac{2n+1}{2n-2}.
% \end{equation}
This indicates that the scaling of the $R$ is given by 
\begin{equation}
    R = \frac{2n+1}{2n-2} \log l +C,
\end{equation}
where $C$ is a constant.
$C$ is evaluated as follows:
\begin{equation}
\begin{split}
&\lim_{l\to\infty} \left( R - \frac{2n+1}{2n-2} \log{l} \right) 
= \frac{1}{1-n}\lim_{l\to\infty}\log \left(l^{\frac{2n+1}{2}} F_{2n}\right).
\end{split}
\end{equation}
According to Eq.~\eqref{eq:FaFa1s}, the limit of the term in the logarithm function is
\begin{equation}
\label{eq:a:s3-const}
\begin{split}
\lim_{l\to\infty} l^{\frac{2n+1}{2}} F_{2n} &= \frac{\Gamma(2n+1)}{p_2^{2n}} \lim_{l\to\infty} \sqrt{l} F_0 
% &= \frac{\Gamma(2n+1)}{p_2^{2n}} \lim_{l\to\infty} \left( \frac{1}{p_2}  -  \frac{2p_{1}}{p_2} F_1\right)\\
= \frac{\Gamma(2n+1)}{p_2^{2n+1}},
\end{split}
\end{equation}
where $\Gamma$ is the gamma function.
Then
\begin{equation}
 C=   \frac{1}{1-n}\log \left( \frac{\Gamma(2n+1)}{p_2^{2n+1}}\right).
\end{equation}

In short, by substituting~ Eqs.~\eqref{eq:tildet},~\eqref{eq:s3-log-lim} and~\eqref{eq:a:s3-const} into Eq.~\eqref{eq:ryee-s3}, the expression of R\'enyi EE (for $n>1$ and $S>1$) 
in the large $l$ limit
becomes\cite{Sugino:2018ab}
\begin{equation}
s^{(n)} = \frac{3}{2}{\left(1+ \frac{1}{n-1}\right)}  \log{l} + \delta,
\end{equation}
where the constant $\delta$ is
\begin{equation}
\label{eq:ryee-form}
\begin{split}
\delta &= \frac{1}{(n-1)}\bigg( \frac{n}{2}\log{\frac{\pi}{2}} -\frac{3 n}{2}\log{\left( 2 + \frac{1}{\sqrt{S}} \right)} - \\ 
&\hspace{0em}  \log{\Gamma(2n+1)}
 + (2n+1)\log {\big( (n-1) \log{S} \big) } \bigg).
\end{split}
\end{equation}

By comparing with the entanglement entropy of a CFT theory, we propose that the coefficient $\frac{3}{2}$ in the logarithmic term of R\'enyi EE is universal,
% which is similar as a central charge in CFT although the Motzkin is not a CFT.
while the constant term $\delta$ is complex and non-universal.
% =======================

% ************************
\section{Numerical result of EE} 
Quantum Monte Carlo (MC) simulations
have been widely used in calculating
R\'enyi EE in quantum many-body systems in recent years~\cite{Hastings:2010zka, Humeniuk:2012xg, inglisWanglandauMethodCalculating2013, broeckerRenyiEntropiesInteracting2014, luitzImprovingEntanglementThermodynamic2014, peiRenyiEntanglementEntropy2014, wangRenyiEntanglementEntropy2014, assaadStableQuantumMonte2015, DEmidio:2019usm, Zhao:2021njg, Yan:2021yzy,wang2024ee}. 
In this section, we use the MC method to estimate the scaling behaviors of VN EE and R\'enyi EE in the Motzkin chain. 

Our simulation is based on the $\text{Swap}$ operator method. 
For pure states that can be written as the product states of two subsystems $A$ and $B$, the $\text{Swap}_{A}$ operator is defined as
\begin{equation}
\begin{split}
&\text{Swap}_{A} \Big( \ket{A_{1}} \otimes \ket{B_{1}}\Big)  \Big( \ket{A_{2}} \otimes \ket{B_{2}}\Big) \\
=  &\Big( \ket{A_{2}} \otimes \ket{B_{1}}\Big)  \Big( \ket{A_{1}} \otimes \ket{B_{2}}\Big).
\end{split}
\end{equation}
The ground state expectation value of the $\text{Swap}_{A}$ operator is the 2nd order R\'enyi EE
% and it can be sampled by Monte Carlo (MC) method
~\cite{Hastings:2010zka, peiRenyiEntanglementEntropy2014, Zhou:2024kcm},
\begin{equation}
\label{eq:ry-swap}
s^{(2)}_A = -\log{\text{Tr}{(\rho_A^2)}} = - \log {\braket{\text{Swap}_{A}}},
\end{equation}
where $\braket{\cdots}$ indicates the expectation value with respect to the replica ground state,
\begin{equation}
 \ket{G}^{2} \equiv \ket{G}\otimes \ket{G}.
\end{equation}

Based on properties of the ground states of colored Motzkin chain, our algorithm can be simplified, 
and then its accuracy significantly improved. 
The details of the algorithm are provided below.

%----------------------------
\subsection{\label{subsec:qmcalg} The Monte Carlo algorithm}
The MC algorithm used to estimate the EE is presented blow.
The ground state given in Eq.~\eqref{eq:groundstate} is positive-definite, hence, 
the expectation value of the $\text{Swap}_A$ operator in Eq.~\eqref{eq:ry-swap} can be written as
\begin{equation}
\label{eq:swap2}
 \braket{\text{Swap}_{A}} = \sum_{i_1,i_2} \frac{1}{N} \bra{G}^2 \text{Swap}_A \ket{M_{i_1}} \otimes \ket{M_{i_2}}.
\end{equation}
Based on this formula, it is 
straightforward to setup a MC algorithm to calculate the 2nd order R\'enyi EE $s^{(2)}$. 
One can simulate two replicas of configurations $M_{i_1}$ and $M_{i_1}$ with equal weights and measure the value of the estimator
\begin{equation}
\label{eq:swap2-estimator}
    \bra{G}^{2} \text{Swap} \ket{M_{i_1}}\otimes \ket{M_{i_2}}.
\end{equation}
However, the symmetries of the colored Motzkin chain ground state allow us to further simplify the algorithm, so that only one replica is needed to sample. 
In this work, we consider the Motzkin chain of the size $2l$, and study the entanglement of the subsystem by cutting at the midpoint.
The length of the subsystem $A$ is
constrained to $l$ in all the discussions below. 

The swappability of Motzkin walks can be defined as follows. 
Exchanging half of the path of walks $M_1$ and $M_2$ generates two new walks denoted as $\tilde{M}_1$ and $\tilde{M}_2$. 
We assert that $M_1$ is swappable to $M_2$ if $\tilde{M}_1$ also qualifies as a Motzkin walk. It is easy to identify the following properties of the Motzkin walks swappability: (1) If $M_1$ is swappable with $M_2$, then $M_2$ is also swappable with $M_1$; (2) If $M_1$ is swappable with $M_2$, and $M_2$ is swappable with $M_3$, it follows that $M_1$ is swappable with $M_3$. 
Consequently, we can conclude that the measurement variable in Eq.~\eqref{eq:swap2-estimator} has only two possible values: It takes the value of one if $M_{i_1}$ and $M_{i_2}$ are swappable; otherwise, it takes the value of zero.

The colored Motzkin walks can then be categorized into classes $c_m$ based on their swappability---walks within each class $c_m$ are swappable with each other. In this context, $m$ represents the height at the midpoint, which corresponds to the number of unpaired steps in half of the Motzkin walks. Notably, there are $S^m$ equivalent classes, where the only distinction is the color permutation of the unpaired steps. Therefore, we define the superclasses $C_m$, which contain the $S^m$ equivalent classes $c_m$.The relationship between the size of the superclass $C_m$ represented as $N_m$, and the size of the class $c_m$ represented as $n_m$, can be succinctly expressed by $N_m = S^m n_m$. With this understanding, we can simulate the distribution of $C_m$. 
The expectation value $\langle \text{Swap}_l \rangle$ can be estimated using the frequency $\Lambda_m$ of each superclass $C_m$. The 2nd order R\'enyi EE is given by the formula:
\begin{equation}
    s^{(2)} = - \log \sum_m S^{-m} \Lambda_m^2.
\end{equation}

To calculate R\'enyi EE of other R\'enyi index, we generalize the $\text{Swap}$ operator and $n$ replicas need to be sampled. However, our algorithm can be directly generalized to simulate the R\'enyi EE of any R\'enyi index $n$, where $n$ could be any positive real number and is not constrained to be an integer, using the following formula:
\begin{equation}
    s^{(n)} = \frac{1}{1-n} \log{ \sum_{m=0}^l S^{-m(n-1)}} \Lambda_m^n.
\end{equation}
Additionally, this algorithm can be employed to determine the VN EE, which is given by:
\begin{equation}
    s^{\text{VN}} = - \sum_{m=0}^l  S^{-m(n-1)} \Lambda_m \log{\Lambda_m}.
\end{equation}

%---------------------
\subsection{\label{subsec:qmcresults}Results of the EE from MC}
\begin{figure}[t]
\centering
\includegraphics[width=.85\linewidth]{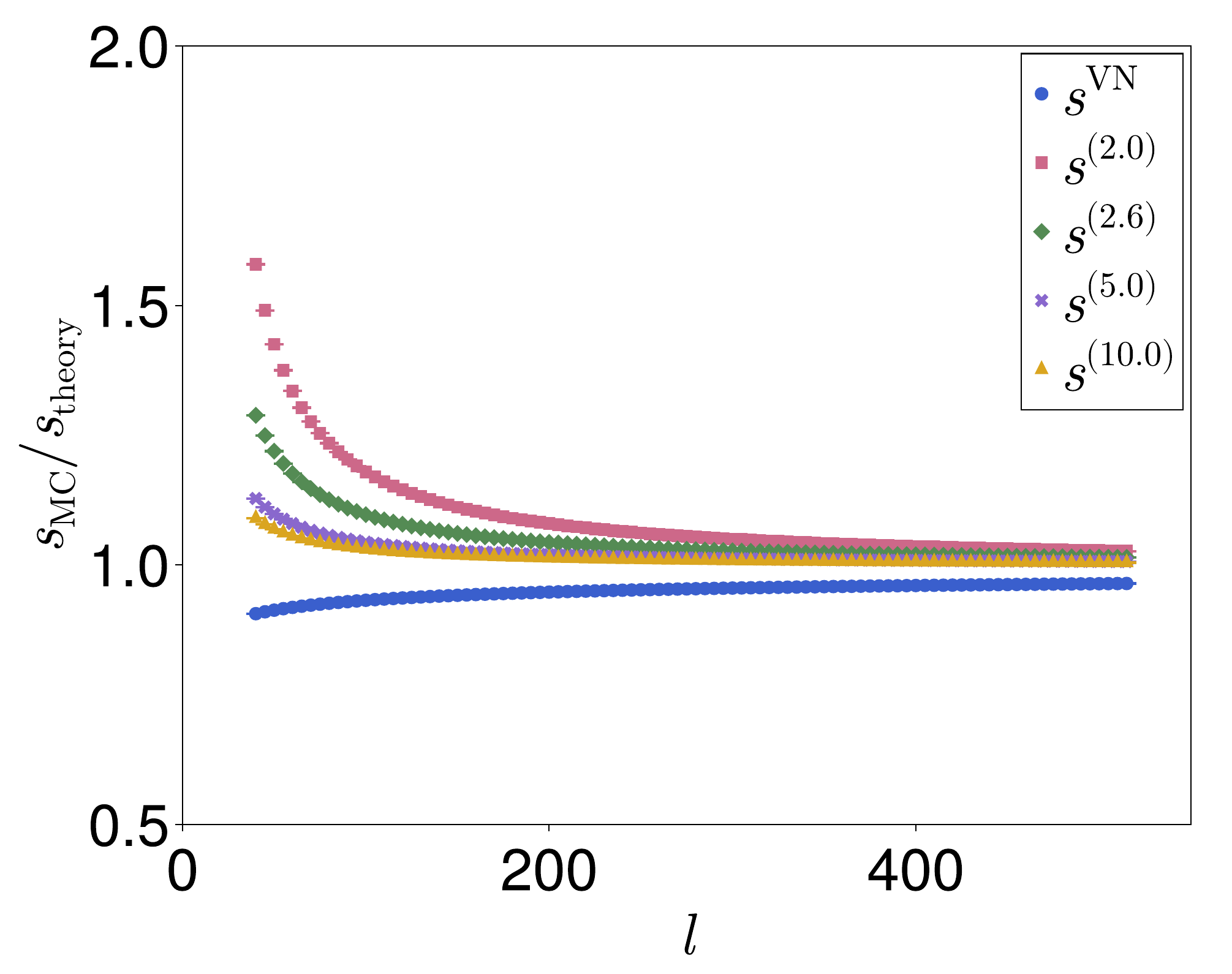}
\caption{\label{fig:s2_n2t} The ratio of the EE results based on MC simulations and analytical calculations.  
The MC are simulated on a two-color ($S=2$) Motzkin chain of size $2l$, and the EE are calculated on the subsystem by cutting at the midpoint. 
As the subsystem size $l$ increases, the ratio tends to be one, showing that the numerical and analytical results agree well for large system sizes.
    }
\end{figure}

The VN EE and R\'enyi EE computed by MC are presented in this subsection. 
Fig.~(\ref{fig:s2_n2t}) shows the comparison of our numerical results and the analytical results, demonstrating a good agreement. 
The ratios converge to $1$ 
as increasing the system size.

%----------------------
%--------------------------

\begin{figure}[tbp]
    \centering
\includegraphics[width=.85\linewidth]{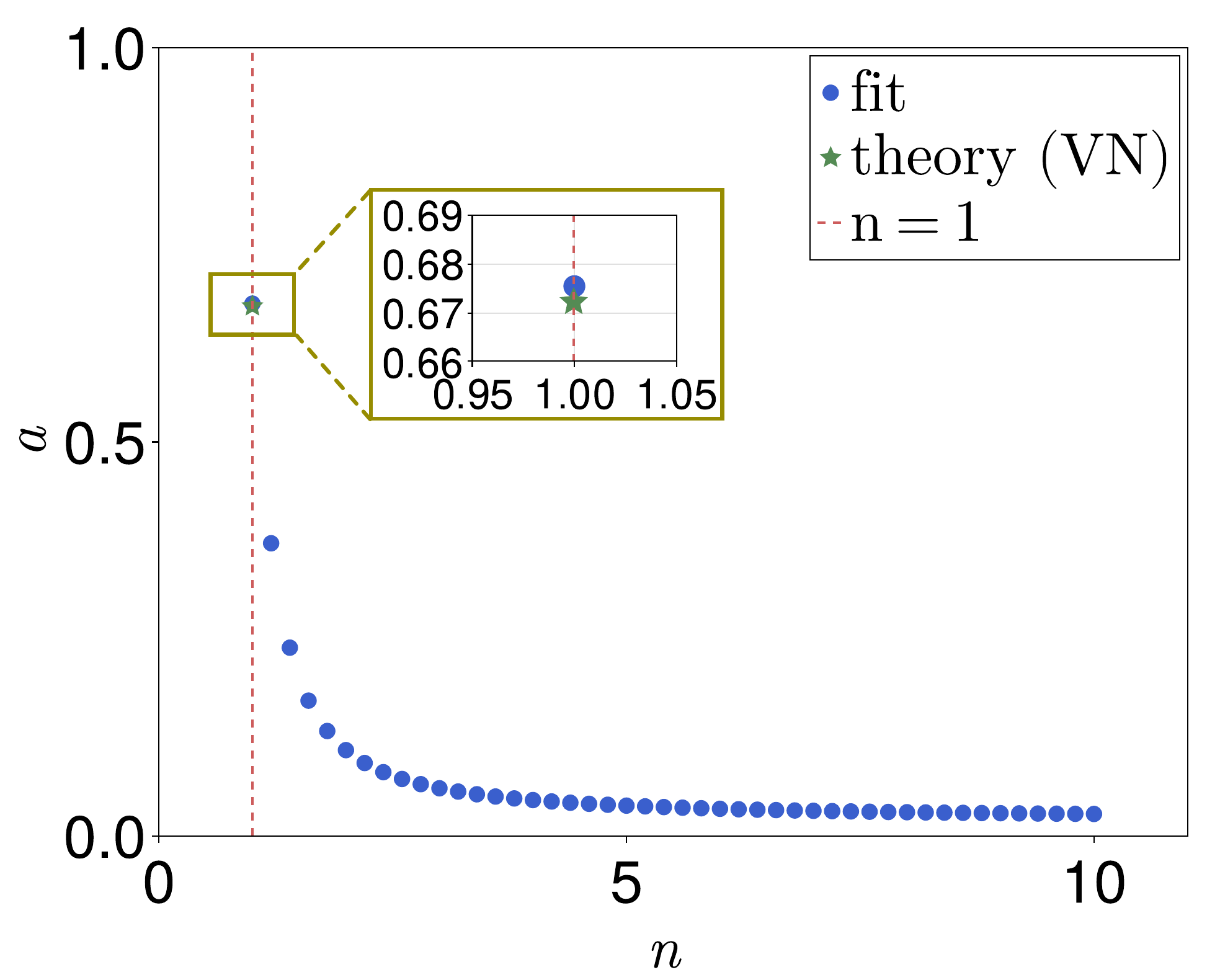}
    \caption{\label{fig:s2_fit_sqrt}The value of the coefficient of the $\sqrt{l}$ term $a$ in EE for a 2-colored ($S=2$) Motzkin chain. The fitting results are illustrated as blue dots, which are fitted with the form: $a \sqrt{l} + b \log{l} +\delta$, where $l$ is the subsystem size. The theory prediction for VN EE is also shown as a green star. The MC are sampled on system size from 80 to 1230 sites.}
\end{figure}

\begin{figure}[tbp]
\centering
\includegraphics[width=.9\linewidth]{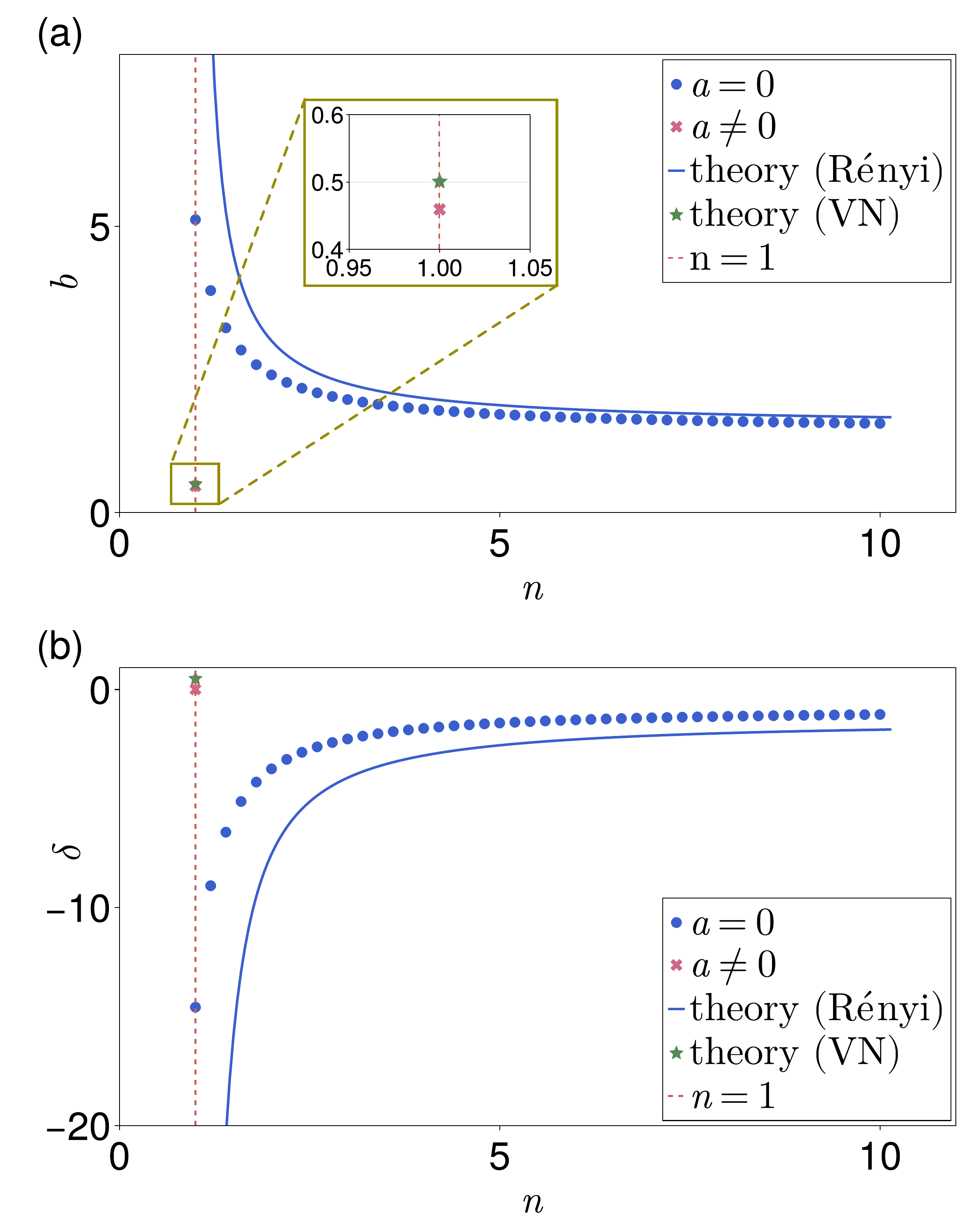}    \caption{\label{fig:s2_fit_ln_const} The coefficient values in the scaling form of R\'enyi and VN EE for the 2-colored Motzkin chain. 
The MC are sampled on the system sizes from 80 to 1230 sites for the both two figures.
The EE data are fitted by the form $b \log{l} +\delta$, represented by the blue dots. 
VN EE data are fitted by the form $a \sqrt{l} + b \log{l} +\delta$, represented by the pink crosses. 
The analytical results are also shown, respectively. 
 ($a$) Value of $b$, the coefficient of the $\log{l}$ term. 
 ($b$) The constant term $\delta$. 
 %\jerry{modified}
 }
\end{figure}

The MC data of the Motzkin chain are analyzed through a fitting process. We fit the data using the form $a \sqrt{l} + b \log{l} +\delta$, 
where $a$ could be zero. 
Here, $l$ represents the size of the subsystem, which is determined by making a cut at the midpoint of the chain with length $2l$. The results of $a$ are presented in Fig. (\ref{fig:s2_fit_sqrt}). 
For the VN EE, indicated by the point at $n=1$, the numerical results closely align with the analytical predictions. Furthermore, our findings indicate that the square root term tends to diminish as $n$ increases, consistent with the previous analysis that the R\'enyi EE does not include a $\sqrt{l}$ term.  
Fig. (\ref{fig:s2_fit_ln_const}) presents the results of $b$ and $\delta$, agreeing with analytical
calculations.
% \zenan{Pls add the system size}.

%------------------------------

%-----------------------
%---------------------------

We observe that there is an EE singularity of the R\'enyi index at $n=1$, as indicated by both analytical and numerical results. The transition from a $\log{l}$-scaling to a $\sqrt{l}$-scaling is governed by the second term in the expansion of the R\'enyi entanglement entropy, as outlined in Eq.~\eqref{eq:ryee-02}:
\begin{equation}
\label{eq:ee-pt}
\begin{split}
  &\frac{1}{1-n} \log{F_{2n}(p_1, p_2) \tilde{T}^{-n}} \\
  = &\frac{1}{1-n} \log{F_{2n}(p_1, p_2) F_2^{-n}(p_1, 0)}.
\end{split}
\end{equation}
The exact value of $F_{2n}(p_1, p_2) F_2^{-n}(p_1, 0)$ can be determined in specific values of $n$ and $S$. For the colorless Motzkin chain, $p_2=0$.
It can be proved that $F_{2n}(p_1, p_2) F_2^{-n}(p_1, 0)>1$ and has an upper bound in the large $l$ limit, 
thus Eq.~\eqref{eq:ee-pt} contributes a constant term to the scaling of EE. For the case $S>1$, however, the value of $F_{2n}(p_1, p_2) F_2^{-n}(p_1, 0)$ can be determined in specific values of $(n-1)\sqrt{l}$ in the large $l$ limit. It can be summarized as
\begin{equation}
  F_{2n}\tilde{T}^{-n} = \begin{cases}
     0, & (n-1)\sqrt{l} \to +\infty \\
    1, & (n-1)\sqrt{l} \to 0^+ \,.
    \end{cases}
\end{equation}
When analyzing the scaling of the R\'enyi EE, we will consider the limit $(n-1)\sqrt{l} \to \infty$. However, this term is treated as zero for the VN EE case. It reflects that the limits of $l\to \infty$ and $n \to 1$ do not commute in the EE calculation. 

As $n = 1$ (the VN EE) is the EE singularity of an infinite Motzkin spin chain, the finite-size modification becomes significantly large near $n = 1$. This finite-size effect is dominated by the term $\frac{1}{(n-1)\sqrt{l}}$, which accounts for the considerable difference observed between the analytical results and the numerical results shown in Fig.~(\ref{fig:s2_fit_ln_const}) as $n$ approaches one. 

\begin{figure}[tbp]
\centering
\includegraphics[width=.9\linewidth]{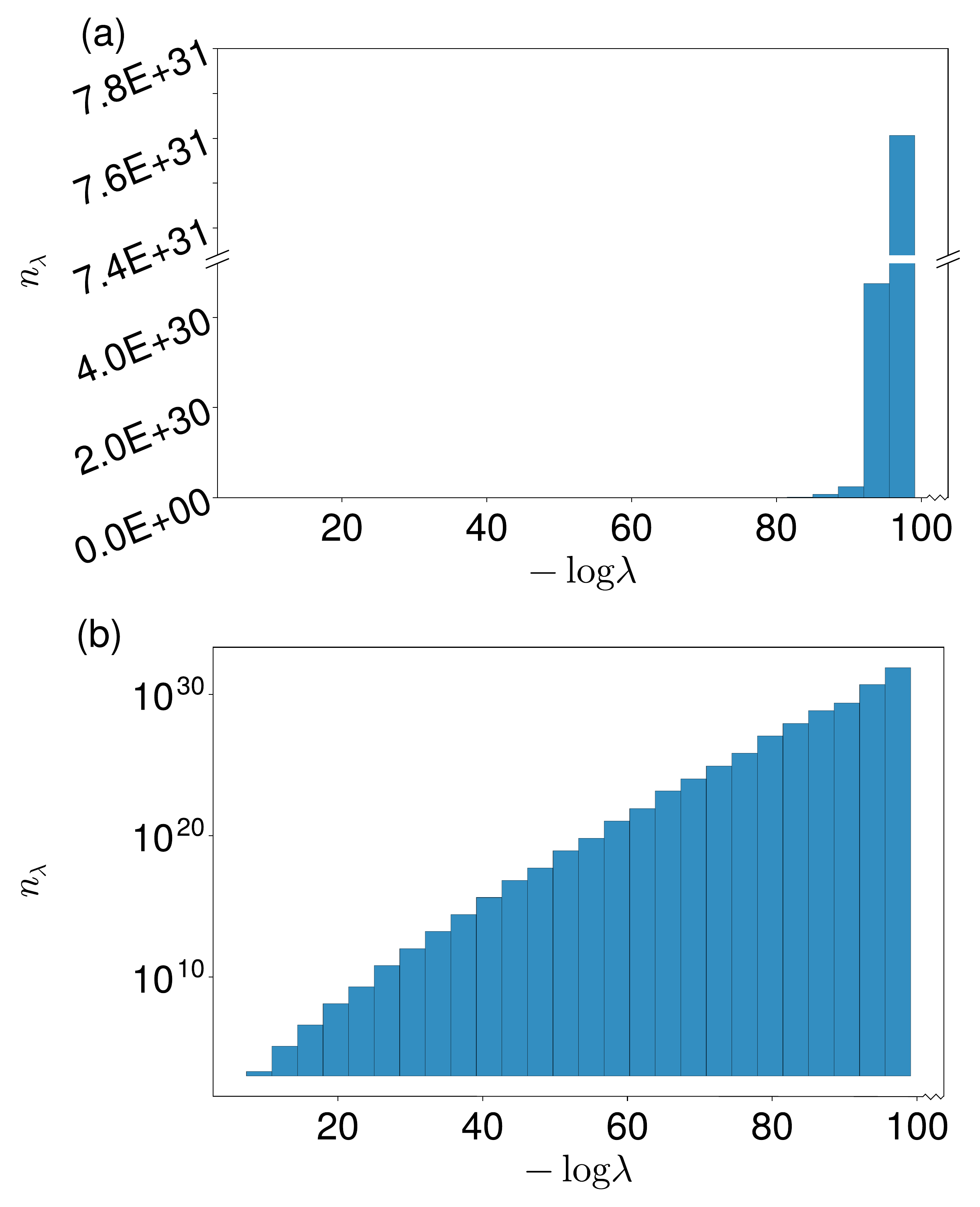}
    \caption{\label{fig:entg_spec} The density of states in the entanglement spectrum, obtained by the reduced density matrix of a spin-2 chain of length 1070. The horizontal axis $-\log\lambda$ is the eigenvalue of the entanglement Hamiltonian ($\lambda$ is the eigenvalue of the reduced density matrix); the vertical axis is the density of states in a small window of $-\log\lambda$ (using a natural or logarithmic scale in figures ($a$) and ($b$)).}
\end{figure}

From the perspective of the entanglement spectrum, 
the different scaling behavior of  the von Neumann (VN) entanglement entropy (EE) from the R\'enyi EE arises from the abundance of high excited levels. 
Entanglement entropies can be understood as thermal entropies of an entanglement Hamiltonian $H_E$. The entanglement spectrum is of the eigenvalues of the entanglement Hamiltonian~\cite{yan2023unlocking,liu2024demonstrating,li2024relevant}. 
For any given R\'enyi index, the R\'enyi EE exhibits the same scaling and is solely dependent on the lower levels in the spectrum. 
In contrast, the exponentially large numbers of high excited levels
— as shown in Fig.~(\ref{fig:entg_spec})
— cause the VN EE to demonstrate a larger scaling behavior than the R\'enyi EE.

%--------------------
%-----------------------

%%%%%%%%%%%%%%%%%%%%%%%%%%%%%%%
\section{Disorder operator} 
The disorder operator\cite{Kadanoff:1970kz,Wegner1971DualityGeneralizedIsing,Fradkin:1978th,Nussinov2009SymmetryPrincipleTopological,Kapustin2014CouplingQFTTQFT,Gaiotto2015GeneralizedGlobalSymmetries,Fradkin:2016ksx,Wen2017ColloquiumZooQuantumtopological,Wen2019ChoreographedEntanglementDances,Ji:2019jhk} is a non-local operator capable of detecting higher-form symmetries, which are often challenging to measure by using local operators. 
For a system with global symmetry, the DO is defined as the symmetry operator that acts on a specific subsystem.
Recent research\cite{Witczak-Krempa:2018mqx,Kong2020AlgebraicHigherSymmetry,Wu:2020yxa,Zhao:2020vdn,Wang:2021lmb,Jiang:2022tmb} has demonstrated that the DO exhibits universal scaling behavior in various quantum systems. Typically, the scaling behavior of the minus of logarithm of DO is similar to the scaling behavior of entanglement entropy, which captures the essential information from CFT.

The Hamiltonian and the ground state of the colored Motzkin chain possess multiple symmetries\cite{Menon:2024vic}. Firstly, the colored Motzkin chain has the continuous $U(1)$ symmetry characterized by the charge $Q^{z}=\sum_j^{2l} S^z_j$, where $2l$ is the system size. 
Consequently, we define the DO as
\begin{equation}
  \begin{split}
   X^z(\theta) = \prod_j^l e^{ i \theta S^{z}_j},
  \end{split}
\end{equation}
which acts solely on half of the chain obtained by cutting at the midpoint. The average of DO is then given by
\begin{equation}
  \label{eq:dis-sz}
   \braket{X^{z}(\theta)}  = \sum_{m=0}^l \lambda_m {\left( \sum_{k=1}^{S} e^{ik \theta}\right)}^m.
\end{equation}
% we obtain the DO in the following form:
% \begin{equation}
%   \begin{split} 
%     \braket{X^z(\theta)} &= \frac{1}{T} \sum_{m=0}^l \frac{m^2}{l} e^{-p_{1}(1,S) \frac{m^{2}}{l}} \left(   \frac{\sin S\frac{\theta}{2} }{ S\sin{\frac{\theta}{2}} } e^{i \frac{S+1}{2} \theta} \right)^m, \\
%     T &= \sum_{m=0}^l \frac{m^2}{l} \exp{\left[-p_1(1,S)\frac{m^2}{l}\right]}.
%   \end{split}
% \end{equation}
% To analyze the scaling behavior in the limit as $l\to\infty$, we again utilize the integral approximation. By setting $\xi = \frac{m}{\sqrt{l}}$, 
By substituting Eq.~\eqref{eq:egvl} into Eq.~\eqref{eq:dis-sz}, setting $\xi = \frac{m}{\sqrt{l}}$, and utilizing the integral approximation, we arrive at the following expression of DO:
\begin{equation}
  \begin{split}
    \braket{X^z(\theta)} = \frac{1}{\sqrt{l}\tilde{T}} \int_0^{\infty} & d \xi \sqrt{l} \xi^2 e^{-p_1(1,S) \xi^2} \times\\   
    &\left[  \frac{\sin (\frac{S \theta}{2}) }{ S\sin{\frac{\theta}{2}}} \right]^{\sqrt{l} \xi} e^{-i \frac{S+1}{2} \theta \sqrt{l} \xi},
  \end{split}
\end{equation}
where $\tilde{T}$ is defined in Eq.~\eqref{eq:tildet}. 

Define the following functions:
\begin{equation}
\label{eq:chi-n}
\begin{split}
    \chi_\alpha(S,\theta, l) = \int_0^{\infty} d\xi &\xi^\alpha e^{-p_1(1,S) \xi^2} 
    %\times \\ & 
\left[  \frac{\sin (\frac{S \theta}{2}) }{ S\sin{\frac{\theta}{2}}} \right]^{\sqrt{l} \xi}  e^{i \frac{S+1}{2} \theta \sqrt{l} \xi},
\end{split}
\end{equation}
\begin{equation}
y(S,\theta)=\log
\left[  \frac{\sin (\frac{S \theta}{2}) }{ S\sin{\frac{\theta}{2}}} \right]
+ i \frac{S+1}{2} \theta.
\end{equation}
Then the DO can be expressed as $\braket{X^z(\theta)} = \chi_2$.
It is easy to prove the following relations:
\begin{equation}
\label{eq:disorder-xz-0101}
\frac{d}{d l} \chi_\alpha = \frac{1}{2 \sqrt{l}}y \chi_{\alpha+1},
\end{equation}
\begin{equation}
\label{eq:disorder-xz-0102}
    \chi_\alpha =  -\frac{\sqrt{l}}{\alpha+1} y \chi_{\alpha+1} + 2 p_1(1,S) \chi_{\alpha+2}.
\end{equation}

Based on Eq.~\eqref{eq:disorder-xz-0101}, we have
\begin{equation}
\label{eq:disorder-xz-de12}
\begin{split}
%\chi_{\alpha+1} &= \frac{2\sqrt{l}}{y} \frac{d}{dl} \chi_{\alpha},\\
\chi_{\alpha+2} = \frac{2}{y^2} \frac{d}{dl} \chi_{\alpha} + \frac{4l}{y^2}\frac{d^2}{dl^2} \chi_{\alpha}.
\end{split}
\end{equation}
Substituting it into Eq.~\eqref{eq:disorder-xz-0102}, it yields \begin{equation}
\label{eq:disorder-xz-02-de12}
    \chi_\alpha =  \left( 2p_1 \frac{2}{y^2} - \frac{2l}{\alpha+1}\right) \frac{d}{dl} \chi_\alpha + \frac{8p_1}{y^2} l \frac{d^2}{dl^2} \chi_\alpha.
\end{equation}

Now look at Eq.~\eqref{eq:disorder-xz-0102}. 
There are two possibilities for the large $l$ scaling of $\chi_\alpha$: 
\begin{itemize}
\item $\chi_\alpha$ and  $\sqrt{l}\chi_{\alpha+1}$ have the same scaling order, while $\chi_{\alpha+2}$ is at a smaller order of $l$;
\item $\sqrt{l} \chi_{\alpha+1}$ and $\chi_{\alpha+2}$ have the same scaling order, while $\chi_{n}$ is at a  smaller order of $l$. 
\end{itemize}
Other cases are impossible, which can be seen by writing down the recurrence relation of $\chi_{\alpha+1},\chi_{\alpha+2}$ and $\chi_{\alpha+3}$. 
Then we have the following arguments about the two cases.

For the first case, $\chi_{\alpha+1} \sim \chi_\alpha/\sqrt{l}$. Substitute it into Eq.\eqref{eq:disorder-xz-0101} and we have 
\begin{equation}
\label{eq:case01-de}
    \frac{d}{d l} \chi_\alpha \sim \frac{y}{l} \chi_\alpha.
\end{equation}
Thus we can suppose that the scaling form of $\chi_\alpha$ is
\begin{equation}
\label{eq:case01-form}
    \chi_\alpha \sim l^{\lambda},
\end{equation}
and the value of $\lambda$ can be determined by substituting Eq.~\eqref{eq:case01-form} into Eq.~\eqref{eq:disorder-xz-02-de12}:
\begin{equation}
\begin{split}
l^\lambda = -\frac{2\lambda}{\alpha+1}  l^\lambda +  \frac{8p_{1}}{y^2} \lambda(\lambda+1) l^{\lambda-1},
\end{split}
\end{equation}
which yields $\lambda = -(\alpha +1)/2$. Thus, the scaling form of $\chi_\alpha$ is $\chi_\alpha \sim l^{-\frac{\alpha+1}{2}}$. 

For the second case, $\chi_{\alpha+1} \sim \sqrt{l} \chi_\alpha$. Substitute it into Eq.\eqref{eq:disorder-xz-0101} and we have 
\begin{equation}\label{eq:case02-de}
     \frac{d}{d l} \chi_\alpha \sim y \chi_\alpha.
\end{equation}
So we can suppose that the leading term of the scaling form of $\chi_\alpha$ is
\begin{equation}
     \label{eq:case02-form}
     \chi_\alpha \sim e^{\beta l}l^{\lambda}.
\end{equation}
The value of $\beta$ and $\lambda$ can be determined by substituting Eq.~\eqref{eq:case02-form} into Eq.~\eqref{eq:disorder-xz-02-de12}
\begin{equation}
\begin{split}
    l^\lambda = &\left(-\frac{2}{\alpha+1}\beta + \frac{8p_1}{y^2}\beta^2 \right) l^{\lambda+1} \\
    &+ \left(\frac{4p_1}{y^2}\beta -\frac{2\lambda}{\alpha+1} + \frac{16p_1}{y^2}\beta\lambda \right) l^\lambda,
\end{split}
\end{equation}
which yields
\begin{equation}
\begin{split}
    \beta &= \frac{y^2}{4p_1 (\alpha+1)}\\
    \lambda &= \frac{\alpha}{2}.
\end{split}
\end{equation}
Thus we have $\chi_\alpha \sim e^{\frac{y^2}{4p_1 (\alpha+1)} l} l^{\frac{\alpha}{2}}$.

It is noteworthy that the term $e^{-p_1(1,S) \xi^2}$ significantly suppresses the contribution from large $\xi$ in the integral given in Eq.~\eqref{eq:chi-n}. Moreover, since $\xi^{\alpha}$ is larger than $\xi^{\alpha+1}$ for small $\xi$, we argue that $\chi_\alpha$ has a larger scaling order in $l$ than $\chi_{\alpha+1}$. This supports the validity of the first case in our situation. Additionally, Numerical calculation of Eq.\eqref{eq:chi-n} and MC  results in the following also confirm the first case.
We calculate the limit ${l\to\infty}$ using Eq.~\eqref{eq:disorder-xz-0101} and~\eqref{eq:disorder-xz-0102},
\begin{equation}
\label{eq:disorder-xz}
\begin{split}
    &\lim_{l\to\infty} \frac{-\log|\braket{X^z(\theta)}|}{\log{l}} = -\lim_{l\to\infty} \frac{l\frac{d}{dl} \chi_2}{\chi_2}\\
    = &\lim_{l\to\infty} \frac{-l\frac{1}{2\sqrt{l} } y \chi_3}{-\frac{\sqrt{l}}{3} y \chi_{3} + 2 p_1(1,S) \chi_{4}} = \frac{3}{2}.
\end{split}
\end{equation}

Therefore the scaling behavior of the minus of the logarithm of DO is expressed as:
\begin{equation}
  \label{eq:dz-theory}
  -\log|\braket{X^z(\theta)}| = \frac{3}{2} \log l +
  \mbox{const}.
\end{equation}
Note that the result holds for both colorless and colored cases. 
The universal coefficient of the 
$\mathrm{log}$ term is $\frac{3}{2}$, the same as in the R\'enyi EE.

%--------------------------
\begin{figure}[tbp]
  \centering
\includegraphics[width=.85\linewidth]{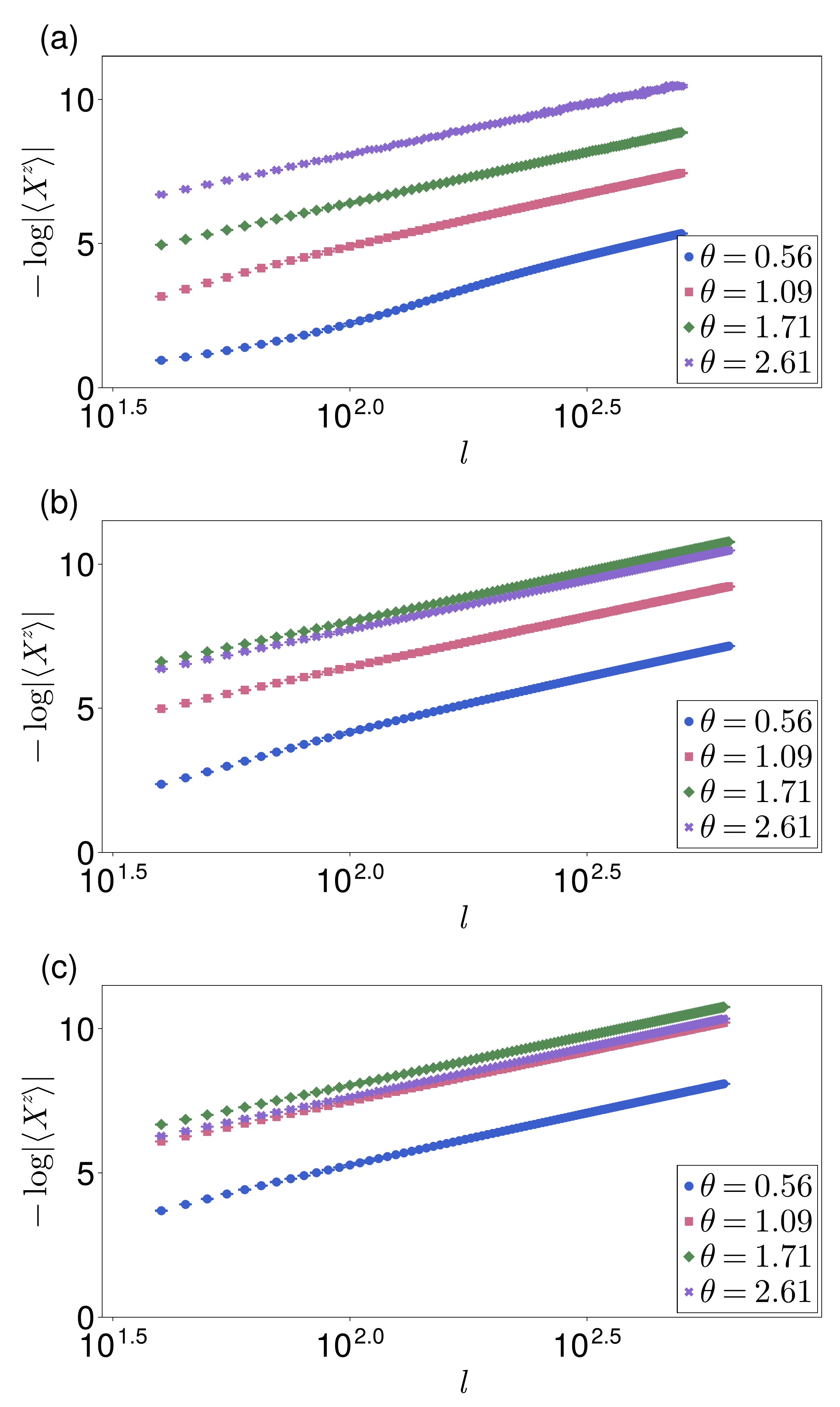}
    \caption{\label{fig:s2_dz_multi} The minus of logarithmic DO under $U(1)$ symmetry with charge $Q^z$. The system is a 2-colored Motzkin chain with size $2l$, and the MC are sampled on system size ranging from 80 to 1230 sites. The horizontal axis is the subsystem size $l$, by cutting at the midpoint of the chain. ($a$) Colorless Motzkin chain. ($b$) 2-colored Motzkin chain. ($c$) 3-colored Motzkin chain.} 
    % {\color{red}the system size?} \jerry{here I don't understand what you mean, the chain length? It is given at the beginning.}
\end{figure}
%----------------------------

We calculate the DO numerically by using MC simulations, and the results of the cases of $S=1,2,3$ are presented in Fig.~(\ref{fig:s2_dz_multi}). 
It shows that the logarithmic scaling agrees well with the analytic result in Eq.~\eqref{eq:dz-theory}.

\begin{figure}[tbp]
  \centering
  \includegraphics[width=.83\linewidth]{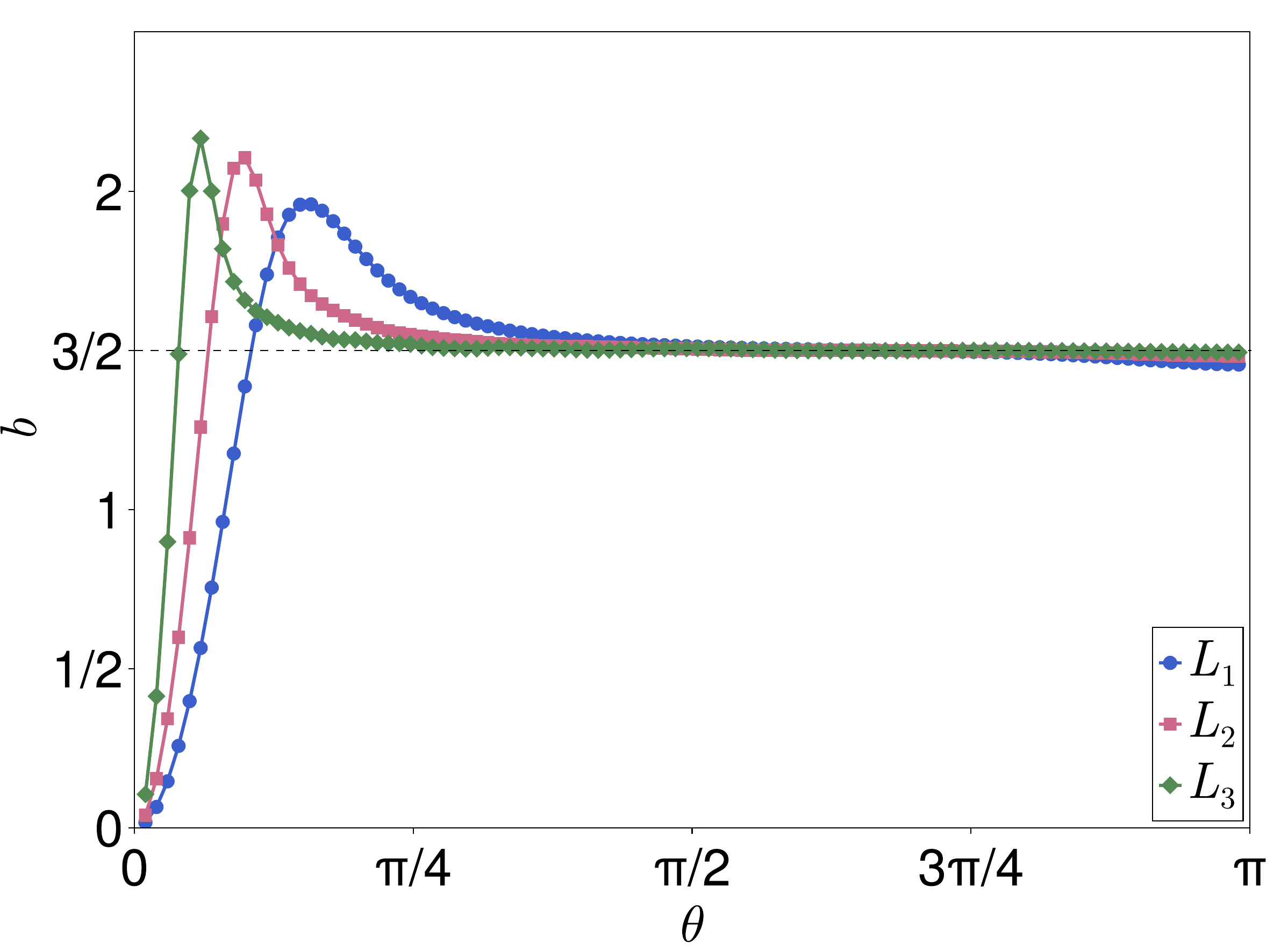}
    \caption{\label{fig:s2-dz-fit-ln} The value of $b$ by fitting the data of $-\log|\braket{X^z(\theta)}|$ within the form $b \log l + const$. The horizontal axis represents the phase $\theta$ in the $U(1)$ symmetry. Different colors correspond to various system size segments. Specifically, $L_1$ are system sizes ranging from 80 to 310 sites, $L_2$ are from 300 to 530 sites and $L_3$ are from 1020 to 1260 sites. The DOs are calculated on subsystems obtained by cutting at the chain at the midpoint.
    The value of $b$ is approximately $3/2$, particularly in the large $\theta$ region. The competition between $\theta$ and $l$ leads to a significant finite-size effect in the small $\theta $ region. This finite-size effect diminishes as the system size $l$ increases, as illustrated in the figure. Therefore, we argue that as the system size increases, $b$ approaches $3/2$ for any value of $\theta$, which aligns with our analytical predictions. }
\end{figure}

The results presented in Eq.~\eqref{eq:dz-theory} hold true in the large $l$ limit. Since $l$ is always paired with $\theta$ in Eq.~\eqref{eq:chi-n}, there is a competition between small $\theta$ values and large $l$ values. This competition leads to significant finite-size effects in the region of small $\theta$. Our simulation also verifies this competition between $\theta$ and $l$. The DOs are fitted using the function $b \log{l} + const.$, based on the data of various segments of system sizes. The results for $b$ are depicted in Fig.~(\ref{fig:s2-dz-fit-ln}).
The figure indicates that for small $\theta$ values, there is a notable difference between $b$ and $3/2$. However, as the system size increases, this difference tends to diminish.
% For small $\theta$, we see a crossover region that the fitted value $b$ is away from $\frac{3}{2}$. This agrees with our previous analyses that the crossover region scale with $l\sim1/\theta^4$.
% \jerry{modified, crossover}

%-----------------------------
\begin{figure}[tbp]
  \centering
  \includegraphics[width=.85\linewidth]{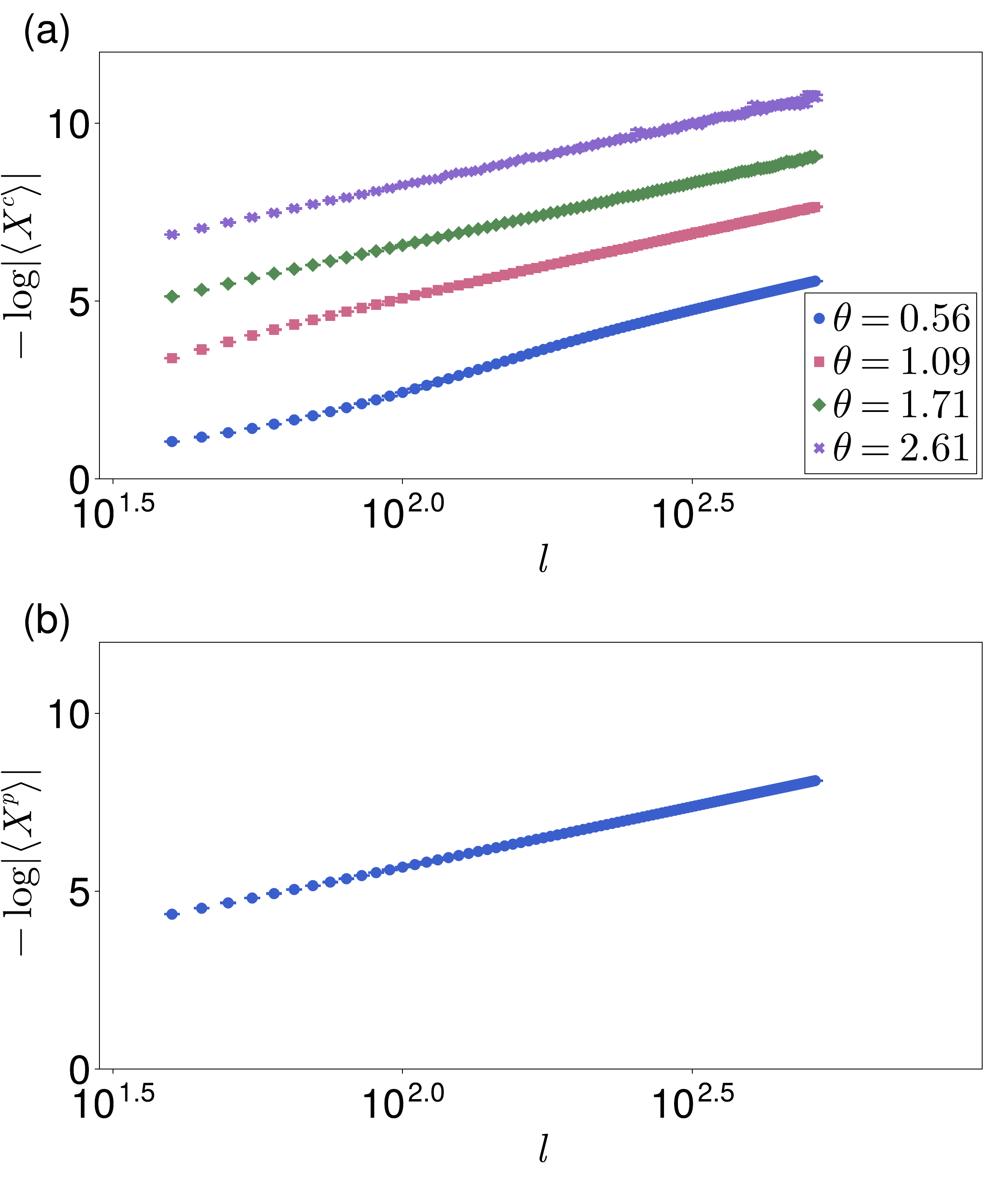}
  \caption{\label{fig:s2-dcdp} 
  The minus of the logarithms of DOs. 
  The system is a 2-colored Motzkin chain, and the MC are sampled on system size ranging from 80 to 1230 sites. 
  The horizontal axis is the subsystem size $l$, by cutting at the midpoint of the chain. 
  ($a$) The DO of $U(1)$ symmetry with charge $Q^c$. 
  ($b$) The DO of permutation symmetry.}
\end{figure}
%------------------------------

In addition to the $U(1)$ symmetry above, the ground state of the Motzkin chain exhibits an additional $U(1)$ symmetry characterized by the charge $Q^c=\sum_j^{2l}\sum_k^S (\ket{\uparrow^k}_j \bra{\uparrow^k}-\ket{\downarrow^k}_j\bra{\downarrow^k})$. 
Then the associated DO is:
\begin{equation}
  \label{eq:dis-c}
  \begin{split}   
    X^{c}(\theta) &= \braket{\prod_j^l e^{\sum_k^S i \theta \left(\ket{\uparrow^k}_j \bra{\uparrow^k}-\ket{\downarrow^k}_j\bra{\downarrow^k}\right)}} \\
             &= \frac{1}{T} \sum_{m=0}^l \frac{m^2}{l} e^{-p_{1}(1,S) \frac{m^{2}}{l}}  e^{ikm \theta}, 
  \end{split}   
\end{equation}
Based on the similar calculation to $X^z$, in the limit of $l \to \infty$, the minus of logarithm of DO also has the $\log l$ scaling behavior,
\begin{equation}
  \label{eq:disorder-xc}
    -\log|\braket{X^c(\theta)}| = \frac{3}{2} \log l +\text{const}.
\end{equation}
The numerical results are illustrated in Fig.~\ref{fig:s2-dcdp}.($a$).

\begin{figure}[tbp]
  \centering
  \includegraphics[width=.8\linewidth]{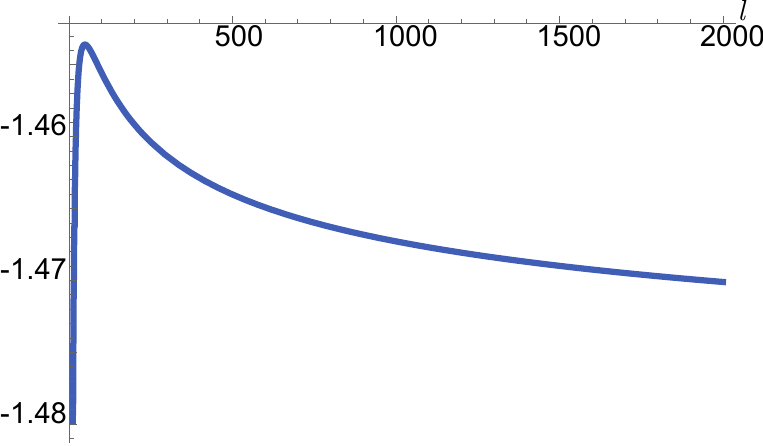}
  \caption{\label{fig:lambda0} The numerical results of the value of $\log{\left((2\sqrt{S}+1)^{l}\sum_{i=0}^{l/2} S^i \frac{l!}{(i+1)!i!(l-2i)!}\right)}/\log{l}$. The horizontal axis is $l$. Based on this calculation results, we argue that as $l\to\infty$, the value goes to $-1.5$.}
\end{figure}
The model has a permutation symmetry concerning different colors. Take the example in Fig. (\ref{fig:walk-example}) for instance, it is still a Motzkin walk after exchanging the blue and pink steps. 
We introduce a DO of $X^p$ that permutes colors within the subsystems. 
It can be proved that the expectation value of this DO equals $\lambda_0$ as outlined in Eq.~\eqref{eq:egvl}, which can be estimated in the limit of large $l$ as\cite{RM2016, Menon:2024vic}:

\begin{equation}
\label{eq:disorder-xp}
\begin{split}
  \braket{X^p} & =  \lambda_0 \\
  &= \frac{l^2}{{\left( 2\sqrt{S} +1\right)}^{2l}T}{\left(\sum_{i=0}^{l/2} S^i \frac{l!}{(i+1)!i!(l-2i)!} \right)}^2\\
  &= \frac{l^{3/2}}{{\left( 2\sqrt{S} +1\right)}^{2l}}{\left(\sum_{i=0}^{l/2} S^i \frac{l!}{(i+1)!i!(l-2i)!} \right)}^2,
  % \\
  % &= \frac{l^2}{\left( 2\sqrt{S} +1\right)^{2l}T} \left( {_2 F_1}(-\frac{l}{2}, \frac{1}{2}-\frac{l}{2}, 2, 4S)\right)^2 \\
  % &= l^{3/2} \, \left({_2 F_1}(-l, \frac{3}{2}, 3, \frac{4\sqrt{S}}{2\sqrt{S}+1})\right)^2,
\end{split}
\end{equation}
% where $_2F_1$ is the hypergeometric function and 
where the last line uses the relation $T=\sqrt{l}\tilde{T}\sim \sqrt{l}$ obtained by Eq.~\eqref{eq:ryee-ttilde} and~\eqref{eq:tildet}. 
We calculate the value of the summation in the bracket numerically due to the difficulty in theoretical analysis, which is shown in Fig.~\ref{fig:lambda0}.
Based on the numerical results, we argue that in the large $l$ limit $\braket{X^p} \sim l^{-3/2}$. Note that we cannot exactly prove the power is strictly $-3/2$, but it looks the same as the previous two DO cases.
Consequently, we believe the minus of logarithmic DO also exhibits logarithmic scaling expressed as:
\begin{equation}
    -\log |\braket{X^p}| = \frac{3}{2}\log{l} +const.
\end{equation}
We simulate this DO by MC. The results also support these findings and have been presented in Fig.~\ref{fig:s2-dcdp}.($b$). 

%We investigate the scaling behavior of three different disorder operators above, based on different symmetries of the ground states in both the colorless and colored cases. Our analysis and numerical results indicate that the scaling of the DOs follow a logarithmic form. 
%Furthermore, we observe that all the coefficients preceding $\log{l}$ are consistently equal to $3/2$ under different symmetries, regardless of whether it is colored or colorless. This universal constant also appears in the R\'enyi EE of the colored case in Eq.~\eqref{eq:ryee-form}. 

%\jerry{more discussion needed. This constant is also the spin-spin correlation, which is $\propto l^{-3/2}$}
% ************************************************************
\section{Summary and discussion} 
It has always been believed that the von Neumann entanglement entropy could be obtained through the analytic continuation of the R\'enyi index from R\'enyi entropy. However, the colored Motzkin chain serves as a counter-example, as the expression for the leading term of R\'enyi entanglement entropy diverges at $n=1$ and an extra term is needed for correction. 
We investigate the scaling behaviors of EE theoretically and numerically to demonstrate the failure of analytic continuation. Mathematically, this singularity is because the limits $l\rightarrow \infty$ and $n\rightarrow 1$ can not commute in the derivation procedures of von Neumann and R\'enyi entanglement entropies respectively. On the other hand, it also could be understood by the exponential increasing density of states in its entanglement spectrum.

We have also explored the scaling of the logarithms of disorder operators under different symmetries in the colored Motzkin chains analytically and numerically. 
The analytic analysis predicts that all the logarithms of DOs exhibit the same scaling as R\'enyi EE. 
Our numerical simulations also confirm analytic calculations.
The scaling of the von Neumann EE,R\'enyi EE and the logarithms of DOs point out that the coefficient of the $\log l$ term is a universal fingerprint to the physics of Motzkin walks.

Although the entanglement entropy is difficult to measure in experiments, even in a small system~\cite{islam2015measuring,PhysRevLett.109.020504}, the disorder operator is an observable that can be easily extracted particularly in cold atom platforms. We propose to probe the constrained physics of Motzkin systems intrinsically via disorder operators experimentally.

\section{Acknowledgment}
We thank the help of Meng Cheng on the analytical calculation. We thank the helpful discussions with Zi Hong Liu. 
JW and CW are supported by the National Natural Science Foundation of China under
the Grant No. 12234016 and No. 12174317.
ZL and ZY acknowledge the start-up funding of Westlake University, the China Postdoctoral Science Foundation under Grants No.2024M762935 and NSFC Special Fund for Theoretical Physics under Grants No.12447119. The authors thank the high-performance computing center of Westlake University and the Beijing PARATERA Tech Co.,Ltd. for providing HPC resources.
This work has been supported by the New
Cornerstone Science Foundation.

\bibliography{ref}
\end{document}